# Investigation of the spectral triplet in strongly coupled quantum dot-nanocavity system

Yasutomo Ota[a,b], Naoto Kumagai[a], Shunsuke Ohkouchi[a,c], Masayuki Shirane[a,c], Masahiro Nomura[a], Satomi Ishida[a], Satoshi Iwamoto[a,b], Shinichi Yorozu[a,c] and Yasuhiko Arakawa[a,b]

a) Institute for Nano Quantum Information Electronics, University of Tokyo, 4-6-1 Komaba, Meguro-ku, Tokyo 153-8505, Japan

b) Institute of Industrial Science, University of Tokyo, 4-6-1 Komaba, Meguro-ku, Tokyo 153-8904, Japan

c) NEC, Miyukigaoka Tsukuba-shi, Ibaraki 305-8501, Japan

We experimentally investigated the excitation power dependence of a strongly coupled quantum dot (QD)-photonic crystal nanocavity system by photoluminescence (PL) measurements. At a low-excitation power regime, we observed vacuum Rabi doublet emission at QD-cavity resonance condition. With increasing excitation power, in addition to the doublet, a third emission peak appeared. This observed spectral change is unexpected from conventional atomic cavity quantum electrodynamics. The observations can be attributed to featured pumping processes in the semiconductor QD-cavity system.

Solid-state cavity quantum electrodynamics (QED) based on semiconductor quantum dots (QDs) has been intensively studied as a key tool for quantum information processing[1-3]. In these studies, single QDs are often considered as atomic two-level systems[4]. However, recent experiments on coupled QD-cavity systems[5-8] have reported several strange phenomena unexpected from conventional atomic cavity QED. One of the major oddities is the so-called non-resonant coupling, which describes strong photon feeding to the cavity mode from the QDs with large spectral detuning from the cavity resonance. Another peculiarity is triplet emission in the strong coupling regime at the resonance condition where vacuum Rabi doublet emission is expected. Both peculiar observations were first reported by Hennessy *et al*[5], and much effort has been made to understand the observations.

With regard to the mystery of non-resonant coupling, several groups have been investigating the mechanisms both experimentally[6,7] and theoretically[9,10]. In contrast, with regard to the spectral triplet, detailed studies have not been conducted so far, and there is little knowledge of the mechanism. The effect of pumping processes on strongly coupled QD-cavity systems is also known little even though photoluminescence (PL) measurements, in which collective carriers are injected around and inside the QDs, are a major experimental tool in semiconductor cavity QED. A theoretical model[11,12] considering incoherent pumping on both the QD and the cavity mode has recently been introduced and applied to explain the pumping power dependence of the vacuum Rabi doublet emission[13]; however, the spectral triplet was outside its scope. Deeper understanding of the peculiar observations is necessary for developing QD-based cavity QED systems for wide application in quantum information technology.

In this paper, we studied the excitation power dependence of a strongly coupled QD-cavity system in the resonance condition by micro-PL measurements. With increasing excitation power, a transition from vacuum Rabi doublet to triplet emission was observed. Quantum correlations of the emitted photons were also investigated and the degradation of the quantum nature along with the increment of the third emission peak were observed. The spectral triplet is attributed to featured pumping processes in semiconductor cavity QED systems, including collective carrier injection inside the host material and incoherent cavity photon pumping by background oscillators.

The investigated sample was grown on a (100)-oriented GaAs substrate by molecular beam epitaxy. First, InAs QDs were grown on an 80-nm-thick GaAs layer on top of a 700-nm-thick $Al_{0.6}Ga_{043}As$ sacrificial layer. The QD density was $\sim 4.0 \times 10^8$ cm$^{-2}$. Then a partially covered island growth technique was applied to the QDs to obtain blue-shifted emission[14]. Finally, the layer of the QDs was buried in the middle of 160-nm-thick GaAs slab layer.

A two-dimensional air-bridged photonic crystal (PhC) was fabricated by forming a triangular array of air holes using a combination of electron beam lithography and etching processes. The lattice constant was $a = 240$ nm and the radius was $r = 0.26a$. The cavity design employed here is a symmetrically modified H1, which is an extension of our previous work[15], and is shown in Fig.1(a). The radius of the nearest-neighbor holes around the defect were shrunk to $0.23a$ and shifted outward by $s_1 = 0.14a$. The second and third nearest-neighbor holes along the ΓK direction and the first nearest holes along the ΓM direction were also shifted by $s_2 = 0.025a$, $s_3 = 0.12a$, and $s_4 = 0.03a$, respectively. The average number of the QDs inside the cavity is estimated to be 0.4.

Micro-PL measurements were performed at 3.3 K with a temperature-controlled liquid helium cryostat. The excitation source was a pulsed Ti:sapphire laser (80 MHz, 2 psec duration) oscillating at 848 nm, where the laser light is predominantly absorbed and creates carriers in the host material (GaAs). The excitation laser was focused onto the sample surface with a spot size of ~3 μm using a microscope objective (40x, N.A. = 0.6). The PL signal was collected using the same objective lens and sent to a 0.75 m grating spectrometer equipped with a cooled CCD (spectral resolution = 0.016 nm) after passing through a half-wave plate and a linear polarizer to effectively detect linearly polarized cavity resonant mode emission. In order to control the wavelength of the cavity mode, we employed the nitrogen gas deposition technique[16]. For photon auto-/cross-correlation measurements, we used a Hanbury-Brown–Twiss setup consisting of a pair of single photon counters located after monochrometers serving as bandpass filters.

Fig.1(b) shows a series of micro-PL spectra of the investigated coupled QD-cavity system at various detuning $\delta$ from −0.3 nm to 0.3 nm. The excitation power was 0.6 μW (measured before the objective lens). Each spectrum is normalized so that its integrated PL intensity is equal to unity. The observed cavity mode (C) originated from one of the fundamental dipole-like modes. The quality factor of the mode was 11400 (116 μeV), measured with a high pumping power of 25 μW at $\delta = -0.6$ nm. The other dipole-like mode split, due to fabrication errors, was at 5 nm shorter than C (not shown). Throughout the cavity scan, the two emission peaks could be seen and exhibited anti-crossing, a signature of strong coupling. The strong coupling behavior is also confirmed by mixing of the linewidths and intensities near resonance. From the smallest splitting, a vacuum Rabi splitting of ~194 μeV was extracted.

Next we studied the pumping power dependence of the strongly coupled QD-cavity system at $\delta \sim 0$. Fig.2(a) shows normalized micro-PL spectra under various pumping powers. At the lowest excitation power of 0.05 μW, the dominant contribution from QD-cavity polariton emission, indicated by blue and red curves, can be seen. With increasing pumping power, an additional emission peak, indicated by the green curve, became prominent and showed dominant emission at a high pumping power of 3 μW. The transition from doublet to triplet emission was not expected from the conventional atomic cavity QED model. This is simply because strongly coupled two-level atom-cavity systems at resonance do not possess their eigenstates at the bare cavity frequency even with incoherent cavity pumping coming from other sources[12].

The three panels in Fig.2(b) show the integrated PL intensities (top), peak positions (middle), and line widths (bottom) of the three peaks of the QD-cavity system at resonance under various pumping powers. These properties were extracted by fitting the spectra with three Lorentzian curves. We assumed a fixed cavity position for fitting the data taken with excitation powers below 0.8 μW. The polariton intensities show linear increase under weak excitation (~0.4 μW) and saturation at higher pumping powers, which is the same behavior as single QDs without strong coupling with a cavity[17]. The intensity of the third emission peak also linearly increases below a power of 0.4 μW and forms an s-shaped curve around the polariton saturation. We consider that this anomalous power dependence does not arise from lasing because of the lack of the characteristic linewidth behavior of a laser. The linewidths of the third peak remain close to the intrinsic linewidth of the cavity, ~116 μeV, at any pumping power. This fact leads us to the same conclusion with Hennessy *et al*[5] that the third emission peak arises from the bare cavity mode. The observed agreement of the polarization between the third emission peak and the bare cavity mode also supports this conclusion. Moreover, the closeness of the measured linewidth values to that of the bare cavity suggests that absorption by oscillators inside the cavity, including the resonant QD, is weakened or off when photon emission from the third peak occurs. Thus, it is appropriate to consider that at that time, the QD is driven to what we call pumped states[18], which are all-inclusive excitonic states other than the cavity-resonant one. The pumped states are formed by capturing or releasing processes in the QD and have weak or no interactions with the cavity mode due to forming dark states or sufficient spectral shifts. Thus, when the QD is in the pumped states, the third peak can appear. We mention the invariance of the peak positions and line widths of the polariton modes against the increment of pumping power, which implies that there are few pump-induced decoherence effects[19] in our sample. We also note that pulsed excitation is not essential. Under continuous-wave laser pumping at the same wavelength, the transition from the doublet to the triplet was also observed in a quite similar manner.

To understand photon sources to the third peak, we measured the quantum correlation between photons emitted from the QD-cavity system. We measured the second-order intensity correlation function for the sum of the three emission peaks $g^{(2)}(t) = \langle I(t)I(t+\tau)\rangle / \langle I(t)\rangle^2$, where $I(t)$ is the emission intensity of the photons at time *t*. In particular, we paid attention to $g^{(2)}[0]$, which is obtained by normalizing the counts in the peak with zero delay ($\tau = 0$) to the averaged counts of the other peaks. An example of the measured coincidence histogram is shown in the inset of Fig.3. The histogram was measured with band-pass filtering centered at 930.2 nm (bandwidth 0.25 nm). The obtained $g^{(2)}[0]$ value is 0.36, which proves the quantum nature of the investigated system. The black balls in Fig.3 show the pumping power dependence of $g^{(2)}[0]$. With increasing pumping power, the $g^{(2)}[0]$ value increases together with the contribution of the third emission peak to the total emission intensity. We analyzed $g^{(2)}[0]$ values in terms of noise contamination to understand the observations. For a single-photon source with noise contribution, we can estimate its $g^{(2)}[0]$ using a simple equation expressed by $1 - (SN/(SN+1))^2$, where *SN* is the ratio of the signal from a single quantum emitter to the noise with Poissonian photon statistics (background emitters)[20]. Here repopulation processes causing multi-time emissions in one pulse are neglected. We assumed that the signal and noise correspond to the total polariton intensity and third peak intensity, respectively. On the basis of this assumption and the fitting results [Fig.2(b) top], we calculated $g^{(2)}[0]$ as a function of the pumping intensity. The estimated $g^{(2)}[0]$ values are shown in Fig.3 as gray triangles. The good agreement indicates that the third emission peak possesses Poissonian photon statistics over a wide range of pumping powers. Thus, it is reasonable to consider that the third emission peak is excited by background oscillators inside the host material, such as deep-level defects in the GaAs[21], tail of the wetting layer[22], and other weakly coupled QDs inside the cavity mode. Note that in PL measurements, care must be taken with respect to the third peak even when PL spectra do not clearly show triplet emission.

Now we discuss the mechanism of the emergence of the third emission peak. In our simple picture, the pumped state of the QD and incoherent pumping of the cavity by background oscillators are essential. As long as the QD can strongly couple to the cavity mode, the system shows vacuum Rabi doublet emission. In contrast, when the QD is driven to the pumped state where its coupling to the cavity is weakened or off, the third peak as the bare cavity mode is excited by background oscillators inside the host material. Integrated spectra of the two emission processes result in the measured spectral triplet. These processes are unique characteristics of semiconductor QD-based cavity QED systems, which allow carriers to be excited collectively inside the host material and the QD. We mention that our theoretical calculations based on the QD model with the pumped state can well reproduce power dependence of the same QD-cavity system, which supports the justification of our interpretations. Details on the calculation will be reported elsewhere.

We also mention that our explanation based on a combination of non(weak)-interacting QD states and cavity pumping is similar to the recent theoretical proposal by Yamaguchi *et al*[23], in which exciton complexes are considered. They considered that the origin of the cavity pumping is the photon feeding from the non-resonantly coupled exciton states instead of the background oscillators.

In summary, we have experimentally investigated a strongly coupled QD-cavity system and observed a transition from doublet to triplet emission along with the increment of the excitation power. The third emission peak is assigned to the bare cavity mode excited by background oscillators with Poissonian photon statistics. We provide a possible explanation for the observations by a combination of the pumped QD state and incoherent cavity pumping, which are essential characteristics of semiconductor systems. We believe that our experimental findings are important for the deep understanding of QD-based cavity QED and will promote its progress.

This work was supported by the Special Coordination Funds for Promoting Science and Technology.

Figures

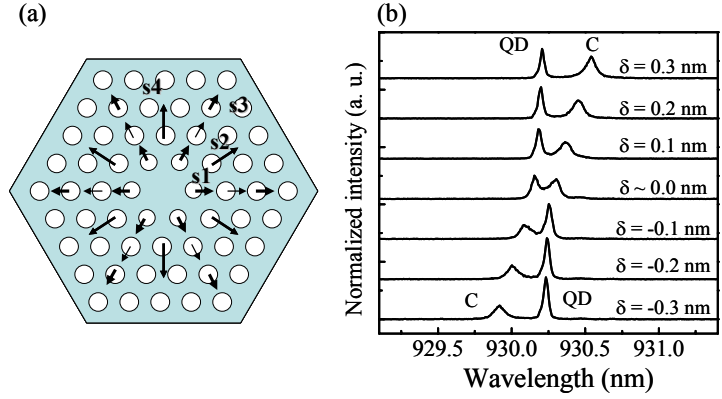

Fig.1. (a) Schematic illustration of the investigated H1-type photonic crystal nanocavity. (b) The observed spectra under various QD-cavity detuning $\delta$ at 3.3 K.

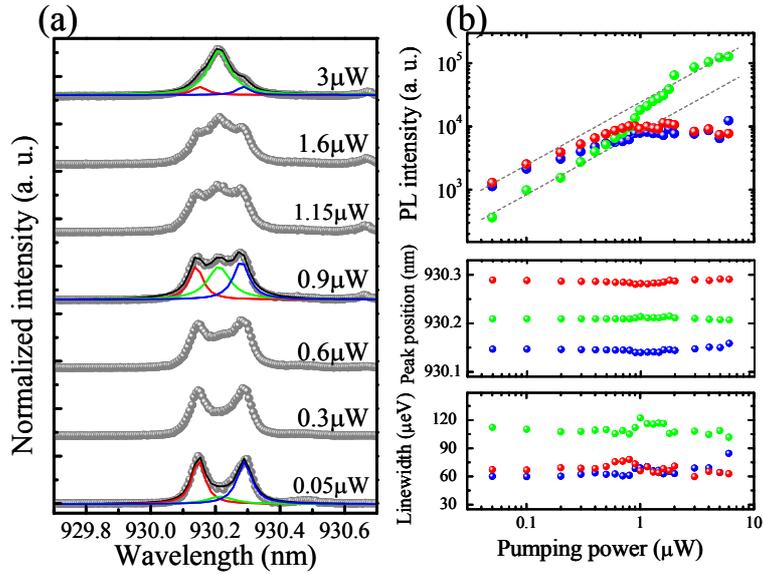

Fig. 2. Excitation power dependence of PL from the strongly coupled QD-cavity system. (a) Excitation power dependence of PL spectra at 3.3 K. Solid lines show the result of fitting by three Lorentzian curves. (b) Properties extracted from the spectra by the fitting: PL intensity (top), peak positions (middle), and line widths (bottom). In both (a) and (b), blue, red, and green lines or points show the fitting results, corresponding to lower and higher polariton and third peak emission, respectively. Dashed gray lines are eye guides for linear power dependence.

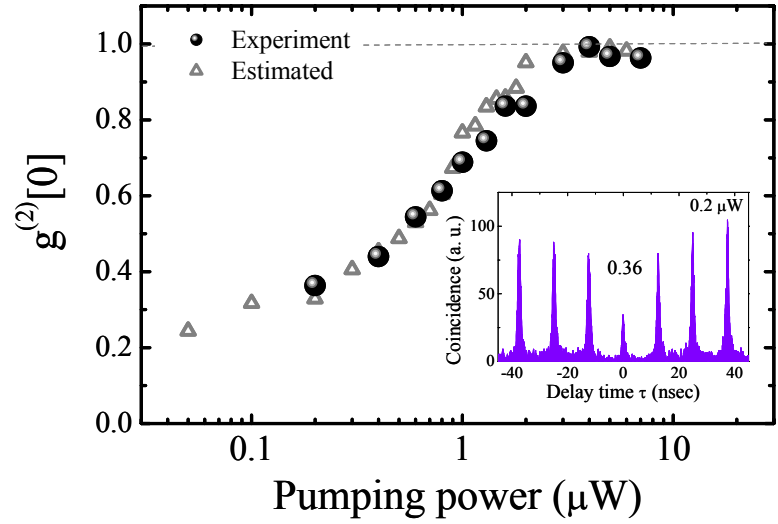

Fig. 3. Measured auto-correlation function at zero-delay time $g^{(2)}[0]$ under various pumping powers. Black balls denote the measured $g^{(2)}[0]$ value from the ensemble emission of the spectral triplet. Gray triangles denote the estimated $g^{(2)}[0]$ values described in the text. Inset shows the obtained coincidence histgram at a pumpimg power of 0.2 μW.